\documentclass[showpacs,showkeys,amsmath,amssymb,pra,twocolumn]{revtex4}

\newcommand{\ie}{\textit{i.e.}}

\usepackage{graphicx}   % Include figure files 
\usepackage{dcolumn}    % Align table columns on decimal point 
\usepackage{bm}         % bolad math 
\newcommand{\rd}{{\rm d}} 
 
\newcommand{\ri}{{\rm i}}

\newcommand{\pmiaou}{p_{\rm miaou}} 

\begin{document}

\title{Scattering tightly bound dimers off a scattering potential}  
 
\author{Christoph Weiss}
\email{Christoph.Weiss@uni-oldenburg.de}

\affiliation{Institut f\"ur Physik, Carl von Ossietzky Universit\"at,
                D-26111 Oldenburg, Germany
}

\keywords{optical lattice, quantum superpositions,
  ultra-cold atoms}
                  
\date{\today}
 
\begin{abstract}
The motion of two attractively interacting atoms in an optical lattice is investigated in the presence of a scattering potential. 
The initial wavefunction can be prepared by using tightly bound exact two-particle eigenfunction for vanishing scattering potential. 
This allows to numerically simulate the dynamics in the generation of two-particle Schr\"odinger cat states using a 
scheme recently proposed for scattering of quantum matter wave solitons.
\end{abstract} 
\pacs{03.65.Nk,03.67.Bg,34.50.-s} 
%checked August 2009

\maketitle

%%%%%%%%%%%%%%%%%%%%%%%%%%%%%%%%%%%%%%%%%%%%%%%%%%%%%%%%%%%%%%%%%%%%%%%%%%%%%%%%

%\newpage

\section{Introduction}

For ultra-cold atoms in an optical lattice~\nocite{LewensteinEtAl07}\cite{LewensteinEtAl07,BlochEtAl08,Yukalov09} dynamical aspects 
include transverse resonances~\cite{MarzlinYukalov05} density waves~\cite{KollathEtAl05}, the evolution of quantum
fluctuations~\cite{FischerEtAl08}, the speed of sound~\cite{BoersEtAl04,LiangEtAl08} and
time-resolved observation and control of superexchange interactions~\cite{TrotzkyEtAl08}. The
aim of the present manuscript is to perform exact two-particle dynamics in an optical lattice
similar to what has been suggested in Ref.~\cite{WeissCastin09},
a bright soliton in a one-dimensional waveguide.  As the dispersion relation for the bound two-particle states in the lattice approach case without lattice for suitable parameters, this can be used to quantitatively test the $N$-particle predictions of Ref.~\cite{WeissCastin09} via exact numerics on the two-particle level for which a soliton is simply a dimer.

Besides the analytical $N$-particle quantum
mechanical calculations~\cite{WeissCastin09}, the scattering of the soliton has also been investigated via 
numerical methods 
on the  $N$-particle level~\cite{Streltsov-2008}. Different approaches to obtain such
Schr\"odinger cat states or related fragmentations have been investigated in Refs.~\cite{StreltsovEtAl09,StreltsovEtAl08}. Contrary to Schr\"odinger cat states of a single atom~\cite{MonroeEtAl96}, cat-like states of
 radiation~\cite{BruneEtAl96} or mesoscopic spin-squeezed states (which have already been
realised experimentally~\cite{EsteveEtAl08}), the experimental realisation of Schr\"odinger
cat states of say, 100 atoms, is still a challenge of fundamental research. Suggestions how interesting quantum superpositions might be
obtained can be found, e.g., in Refs.~\cite{CastinDalibard97,
RuostekoskiEtAl98, CiracEtAl98, DunninghamBurnett01, MicheliEtAl03, MahmudEtAl03,TeichmannWeiss07,Dounas-frazerEtAl07, DagninoEtAl09} and references therein.

For bright  quantum matter wave solitons~\cite{LeeBrand06,CastinHerzog01,CalabreseCaux07}, the mean-field (Gross-Pitaevskii) limit has been shown to be achieved
already for particle numbers as low as $N\gtrapprox 3$~\cite{MazetsKurizki06}. Many of the
papers published after the ground-breaking experiments~\cite{KhaykovichEtAl02,StreckerEtAl02}
solve the Gross-Pitaevskii equation for solitons. However, any
mesoscopic entangled state which involves superpositions of wavefunctions cannot be
described by a non-linear equation and therefore the reasoning of Ref.~\cite{MazetsKurizki06}
is not valid in the situation considered here. Thus, instead of applying the Gross-Pitaevskii equation, the
$N$-particle Schr\"odinger equation has to be used to reveal true quantum behaviour of a
soliton created from a Bose-Einstein condensate. Under experimentally realistic conditions,
the Schr\"odinger equation is given by the analytically solvable Lieb-Liniger(-McGuire)
model. 

The challenge of the generation of mesoscopic superpositions via scattering of solitons is that to add
a scattering potential removes the separability of the centre-of-mass motion and the relative
motion; in order to avoid that the scattering potential acts like a beam splitter on each single atom
(rather than the entire soliton), the initial state has to be prepared carefully. Mesoscopic
entangled states with the soliton being in a quantum superposition with $\approx 50\%$
probability of moving to the right/left should thus be obtainable. The probability to find
-- in a \textit{single measurement}\/ -- (at least) one particle moving to the right and at  (at least)
one particle moving in the other direction will be negligible. However, this will not be
enough to prove that the two parts of the wavefunction really are in a quantum superposition
-- if someone claims that a coin is in a quantum superposition of heads and tails, an
experiment showing only the classical outcomes would hardly convince anyone. The experimental verification
could be delivered via interference experiments~\cite{WeissCastin09}.

Rather than dealing with bright $N$-particle quantum solitons, this paper treats a
simpler but nevertheless instructive case: dimers in an optical lattice. 
The paper is organised as follows: after a short summary of how to describe the scattering of
bright solitons analytically~\cite{WeissCastin09} (Sec.~\ref{sec:liebliniger}), the two-particle bound states used to
describe the scattering of the dimer are introduced in
Sec.~\ref{sec:two}. Section~\ref{sec:results} shows the numeric results in the limit where the motion in the optical lattice mimics the motion without lattice. 

\section{\label{sec:liebliniger}Effective potential approach}

The Hamiltonian of the Lieb-Liniger-McGuire~\cite{LiebLiniger63,McGuire64} model with attractive interaction and
an additional scattering-potential~$V$  is given by
\begin{equation}
\label{eq:H}
\hat{H}=\sum_{i=1}^{N}\frac{p_i^2}{2m} +
  g\sum_{i<j} \delta(x_i-x_j) + {V}\;,\quad g<0\;.
\end{equation}
Bright solitons~\cite{KhaykovichEtAl02} are well described by this model. For $V=0$, exact
eigenfunctions of this Hamiltonian are known. Solutions corresponding to $N$-particle solitons with
momentum~$k$ read:
\begin{equation}
\label{eq:psink}
\psi_{N,k}(\underline{x})=\exp\left(-\widetilde{\beta} \sum_{1\le\nu<\mu\le
    N}\left|x_{\nu}-x_{\mu}\right|+\ri k\sum_{\nu=1}^{N}x_{\nu}\right)\;,
\end{equation}
where 
\begin{equation}
\label{eq:betatilde}
\widetilde{\beta} = -mg_{\rm
  1d}/(2\hbar^2)>0\;.
\end{equation}

The corresponding energies are given by
\begin{equation}
E_k=E_0+\frac{N\hbar^2k^2}{2m}
\end{equation}
where 
\begin{equation}
\label{eq:E0}
E_0 = -\frac{mg^2}{24\hbar^2}N(N^2-1)
\end{equation}
is the ground state energy of the system~\cite{McGuire64}. As long as the kinetic energy is
not too large, these states are separated from the first excited internal state (which
corresponds to one particle having left the soliton) by a finite energy barrier~$\propto
g^2N^2$ (see, e.g., Ref.~\cite{CastinHerzog01}).

Had the scattering potential been a function of the centre of mass of all $N$ particles ($V=V\left({\textstyle\frac1N\sum_{\nu}x_{\nu}}\right)$), the
situation would have been easy
as the centre of mass and
  relative coordinates then still separate.
However, the potential in the Hamiltonian~(\ref{eq:H})
 is given by
\begin{equation}
V=\sum_{\nu}\widetilde{V}\left({x_{\nu}}\right)
\end{equation}
It would nevertheless be tempting to argue that, given the fact that the particles are
tightly bound, they behave essentially as a single particle and one could thus
approximate~$\widetilde{V}\left({x_{\nu}}\right)$ by $\widetilde{V}\left(X\right)$ and thus
\begin{equation}
V\approx N\widetilde{V}(X)
\end{equation} where 
\begin{equation}
X=\frac1N\sum_{\nu=1}^Nx_{\nu}
\end{equation}
is the centre-of-mass coordinate. However, this approximation can give wrong results (as will
be shown towards the end of this paper) and the mathematically justified~\cite{WeissCastin09}
effective potential approach:
\begin{equation} \hat{H}_{\rm eff} = -\frac{\hbar^2}{2M}\partial^2_X +
  {V}_{\rm eff}(X)
\end{equation}
has to be used. The effective potential is given by the convolution~\cite{WeissCastin09}
\begin{eqnarray}\label{eq:Veffend}
V_{\rm eff}(X)=\int\rd^Nx&&\hspace*{-0.6cm}\delta\left({\textstyle
    X\!-\!\frac1N\sum_{\nu=1}^Nx_{\nu}}\right)\\\nonumber
&\times&|C{\psi}_{N,k}(\underline{x})|^2\sum_{\nu=1}^N\widetilde{V}(x_{\nu})\;.
\end{eqnarray}
This approach is valid for sufficiently well behaved potentials (like a laser focus) and for
solitons which cannot break apart for energetic reasons (see the paragraph below Eq.~(\ref{eq:E0})).

\section{\label{sec:two}Two-particle bound states}

Two-particle bound states in optical lattices are interesting both experimentally~\cite{WinklerEtAl06} and
theoretically
\nocite{PiilMolmer07,PetrosyanEtAl07}\nocite{ValientePetrosyan08}\cite{PiilMolmer07,PetrosyanEtAl07,ValientePetrosyan08,WeissBreuer09,JinEtAl09,WinklerEtAl06};
recently even three-particle bound states~\cite{ValienteEtAl09} have been
investigated. Within a Bose-Hubbard Hamiltonian,
\begin{eqnarray}
\label{eq:lattice}
\hat{H}_{\rm lattice} &=&
-J\sum_j\left(\hat{c}^{\dag}_j\hat{c}^{\phantom{\dag}}_{j+1}+\hat{c}^{\dag}_{j+1}\hat{c}^{\phantom{\dag}}_j\right)\\
\nonumber& & 
+ \frac U2\sum_j \hat{n}_j\left(\hat{n}_j-1\right),\quad U<0
\end{eqnarray}
 one can use \textit{exact}\/ eigen-states to do the
numerics (the restriction to negative pair interactions $U$ is not necessary, however the idea is to
discuss a case close to the bright solitons created from attractive Bose-Einstein
condensates; the creation/annihilation operators of particles at lattice cite~$j$ are denoted
by $\hat{c}^{\dag}_j$/$\hat{c}^{\phantom{\dag}}_{j}$). Rather than using the approach via Green's functions of Ref.~\cite{WinklerEtAl06},
we proceeded in Ref.~\cite{WeissBreuer09} along the lines of Ref.~\cite{Weiss06b}  to show in a straight-forward  but somewhat lengthy calculation that the dimer
wavefunctions in a tight-binding lattice with band-width $W$ and lattice spacing~$b$ are given by
\begin{equation}
\label{eq:eigenfunkt}
c_{\nu\mu} = \left\{\begin{array}{ll}
 x_-^{|\mu-\nu|}\exp\left[ikb(\nu+\mu)\right]&:\;\mu\ne\nu\\
\exp\left[ikb(\nu+\mu)\right]/\sqrt{2}&:\; \mu = \nu
\end{array}\right.,
\end{equation}
where  
\begin{equation}
 x_-=\sqrt{\frac{U^2}{16J^2\cos^2(kb)}+1}-\frac{|U|}{|4J|\cos(kb)}\;;
\end{equation}
the coordinates $\nu$ and $\mu$ label the lattice points on which the particles ``sit'' and
as a basis the Fock basis is used (\ie\/ $\nu=\mu$ refers to the Fock state with two
particles at the lattice-site~$\nu$).
The interaction energy of two particles sitting at the same lattice site is $\hbar\kappa<0$;
for the wavefunction to be normalisable for fixed centre of mass, $kb<\pi/2$ (which implies
$|x_-|<1$). To avoid a breaking of the dimer, the energy
\begin{equation}
E_{\rm two} = -4J\sqrt{\frac{U^2}{16J^2}+ \cos^2(kb)}
\end{equation}
should be lower than two times the minimal possible energy of two particles:
\begin{equation}
\label{eq:bedB}
E_{\rm two} < -4J
\end{equation}
As for the single particle energy band,
\begin{equation}
E_{\rm one} = -2J\cos(kb),
\end{equation}
the dispersion relation for two particles approaches the free-particle behaviour in the limit
$bk\to 0$ which
can  thus be used to model particles not restricted by optical potentials:
\begin{equation}
E_{\rm two} \simeq -4J\left(\frac{U^2}{16J^2}+ 1\right)^{1/2} +\frac{8J^2b^2}{16J^2+U^2}k^2\;.
\end{equation}
By choosing the parameters to approximate the dispersion relation for a two-particle
``soliton'' (cf.\ Eq.~(\ref{eq:psink}))
\begin{equation}
\label{eq:zweisoliton}
E_{N=2\;\rm soliton} = -\frac{\hbar^2\widetilde{\beta}^2}m+\frac{\hbar^2k^2}m\;,
\end{equation}
the effective potential approach of Ref.~\cite{WeissCastin09} (cf.\ Eq.~(\ref{eq:Veffend})) will thus be valid for small
enough~$b$.

\section{\label{sec:results}Numeric results}
\begin{figure}
\begin{center}
\includegraphics[width=1.0\linewidth]{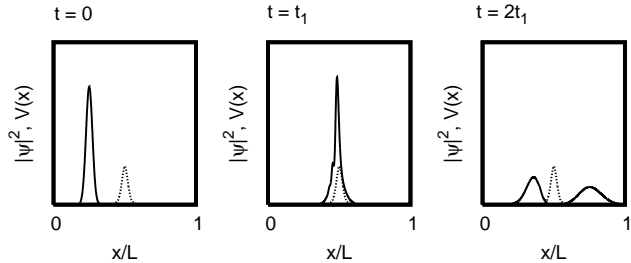}
\end{center}
\caption{A single particle is prepared in a Gaussian wavefunction which initially moves to
  the right. After the particle
  is scattered off a smooth potential (dashed line), the final state, the wavefunction is partially
  moving to the right and partially moving to the left. The data was obtained by numerically solving the
  time-dependent Schr\"odinger equation. \label{fig:refl}}
\end{figure}
The initial centre-of-mass wavefunction in Ref.~\cite{WeissCastin09} is suggested to be a
Gaussian (obtained by preparing the system in a swallow  harmonic oscillator potential).
If the system can be described within a centre-of-mass approximation, the expected result
will always look similar to Fig.~\ref{fig:refl}. For the initial Gaussian to be realisable
experimentally, the length scale on which the wavefunctions of the relative coordinates
decays has to be smaller than the width of the initial wavefunction. Given the fact that the
dimer cannot break into two free particles, this implies that in the
final wavefunction measuring one particle on one side of the barrier would also lead to
measuring the other particle on the same side - thus the final state indeed is a
Schr\"odinger cat state.

\begin{figure}
\begin{center}
\includegraphics[angle=-90,width=0.9\linewidth]{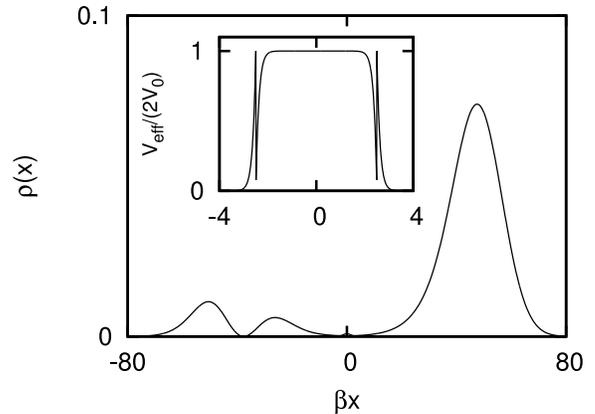}
\end{center}
\caption[three]{\label{fig:three} Particle density~$\varrho(x)$ after a dimer  was scattered off a step potential~(\ref{eq:Vstep}).
The effective potential~(\ref{eq:Veffend}) in the inset explains why resonances can occur. They lead the dimer (which for energetic reasons cannot break into two free particles) to split into three parts rather than the expected two.
The initial wavefunction is given by Eq.~(\ref{eq:initial}) with $a/\widetilde{\beta}=14$, $k_0/\widetilde{\beta}=0.75$ and $x_0\widetilde{\beta}=-50$; $V_0 2m\widetilde{\beta}^2/\hbar^2\simeq0.197$. The calculations were done on a $1601\times 1601$ lattice.
}
\end{figure}
Figure~\ref{fig:three} shows the result for a two-particle soliton scattered off a step
potential,
\begin{equation}
V_{\rm step}= \left\{\begin{array}{ll}
 V_0&:\;\left|x\right|\le\ell\\
0&:\;\rm else
\end{array}\right.\;.
\label{eq:Vstep}
\end{equation}
The initial wavefunction is constructed as a superposition of the eigen-solutions
(\ref{eq:eigenfunkt}) with $0<k<\widetilde{\beta}$ which $\widetilde{\beta}$ ensures the
validity of Eq.~(\ref{eq:bedB}) (cf.\ Eq.~(\ref{eq:zweisoliton})):
\begin{eqnarray}
\label{eq:initial}
\Psi(x_1,x_2;t=0)\propto\int_{0}^{\widetilde{\beta}}\rd k
\exp\left({\textstyle-a^2(k-k_0)^2/2}\right)\\ \nonumber
\times \psi_{2,k}\left(x_1-x_0,x_2-x_0\right)\;.
\end{eqnarray} 
If~$a$ is not too small this leads to a Gaussian centre-of-mass wavefunction initially centred at~$x_0$. 
Surprisingly, for the parameters chosen in Fig.~\ref{fig:three} the wavefunction splits into
three rather than the expected two parts. If one of the particles was measured, the second
would be much closer to the location of the first than the width of the three moving
wave-packets. The strange behaviour of Fig.~\ref{fig:three} clearly demonstrates that the
centre-of-mass approximation is not valid here: the effective potential~(\ref{eq:Veffend}) as
displayed in the inset of  Fig.~\ref{fig:three} at least qualitatively explains why resonances can occur which make parts 
 the dimer  wave-function stay at the potential. 

\begin{figure}
\begin{center}
\includegraphics[angle=-90,width=0.9\linewidth]{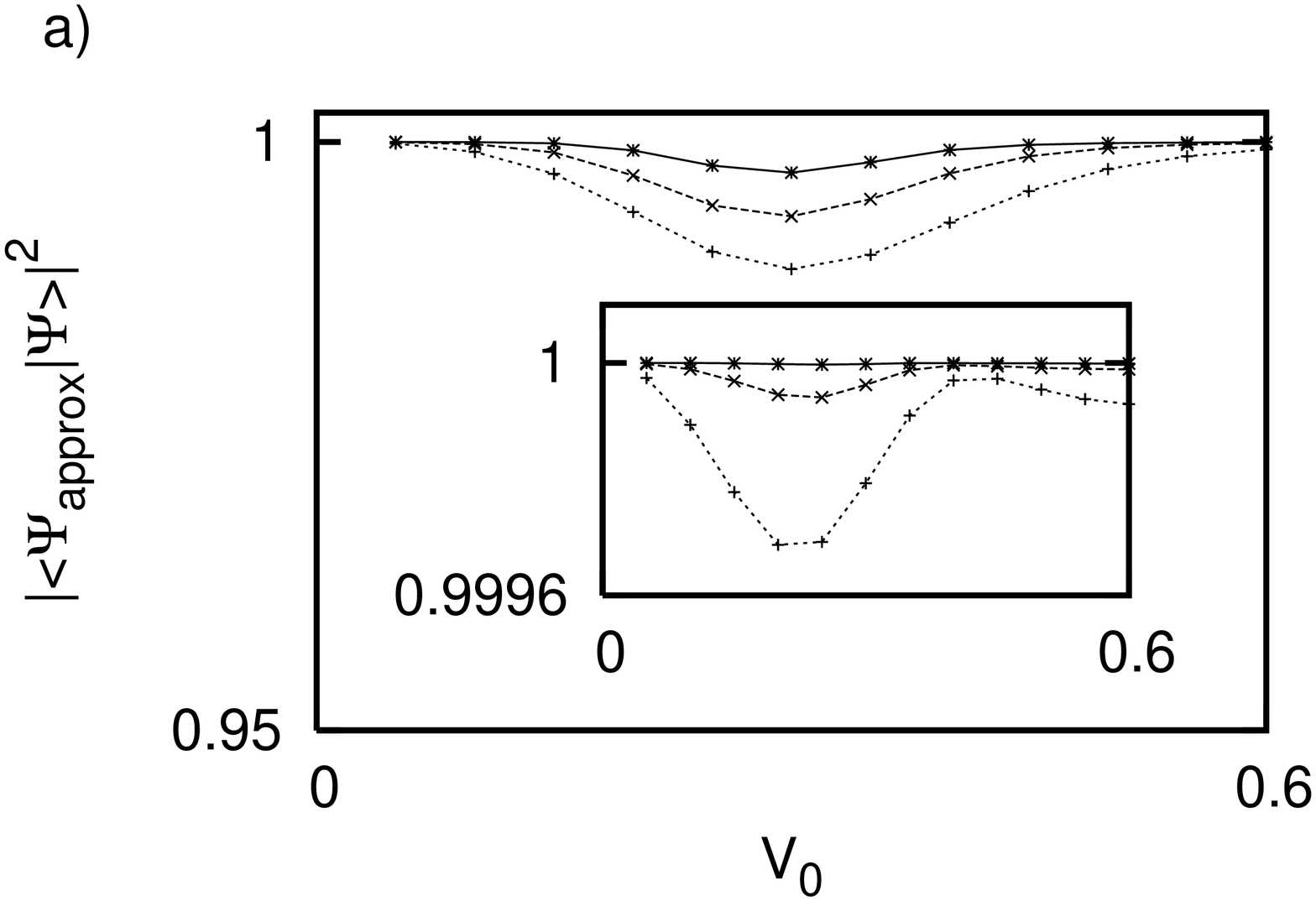}
\includegraphics[angle=-90,width=0.9\linewidth]{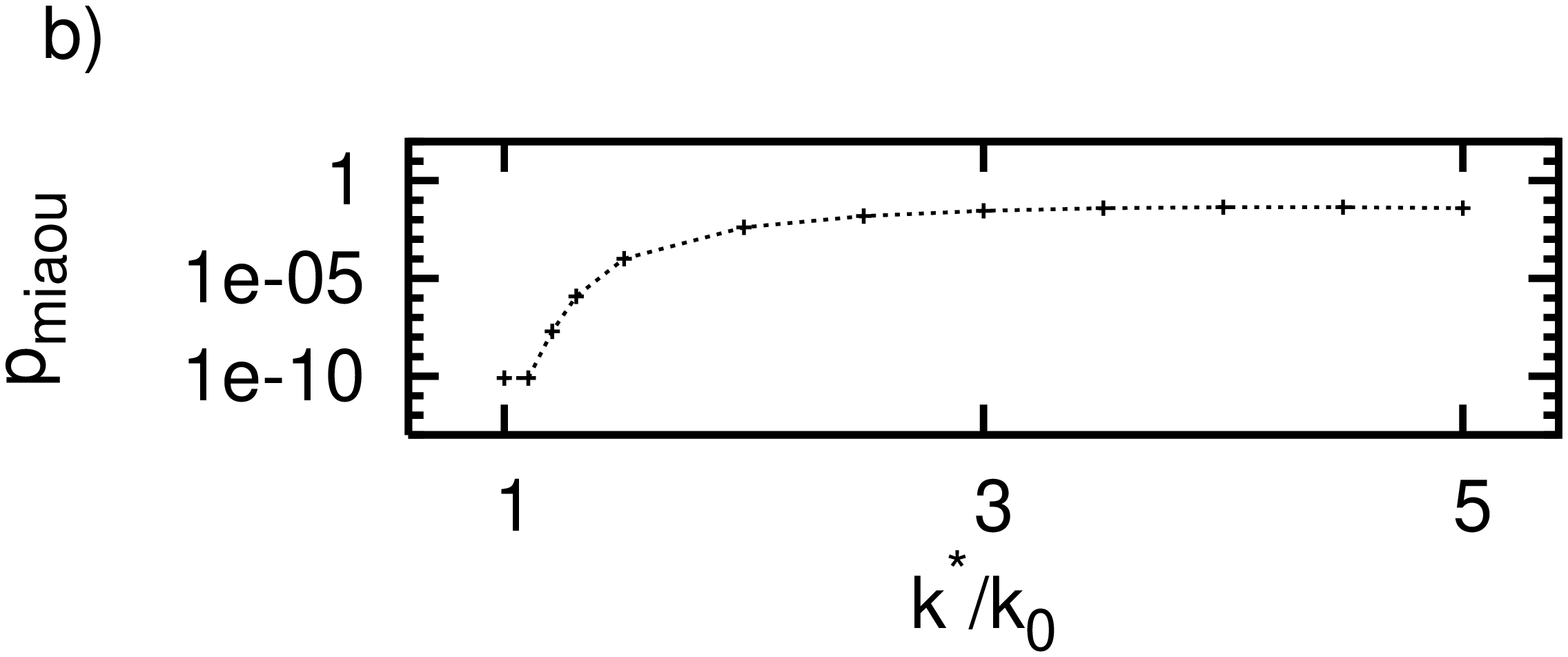}
\end{center}
\caption[four]{\label{fig:miaou}a) Modulus squared of the scalar product after scattering of the solution of the
  approximate Schr\"odinger equation corresponding to Eq.~(\ref{eq:potapprox}) with the
  unapproximated solution (both calculated on a $801\times801$ lattice). The potential is
  given by Eq.~(\ref{eq:potexact}). From top to bottom: $\ell\widetilde{\beta} = 2.5\,,\;
  1.5\,,\; 1$. Inset: with the approximation expressed by Eq.~(\ref{eq:potapproxbetter}), the agreement is much better. The lines are a guide to the eye.
b) If the soliton splits into two parts, $\pmiaou\approx 0$ (Eq.~(\ref{eq:pmiaou}); plotted as a function of the centre of
the momentum distribution) indicates entanglement (for $\ell\widetilde{\beta} = 2.5$). The wave-packet
was defined to be $\propto\int_{0}^{(k^*/k_0)\widetilde{\beta}}d k
\exp\left({\textstyle-a^2(k-k^*)^2/2}\right)\psi_{2,k}\left(x_1-x_0,x_2-x_0\right)$; $V_0$
was chosen such that Eq.~(\ref{eq:potapprox}) predicts $50\!:\!50$ splitting; $k_0/\widetilde{\beta}=0.75$). 
}
\end{figure}

However, rather than being able to qualitatively understand the scattering behaviour, the effective potential approach can also be tested quantitatively for realistic scattering potentials. If the scattering potential is realised via the focus of a laser beam, the potential can be approximated by a Gaussian or even the potential
\begin{equation}
\label{eq:cosh}
 V_{\rm c}(x) = \frac{V_0}{\cosh(x/\ell)^2}
\end{equation}
which is analytically solvable on the single particle level.
Figure~\ref{fig:miaou}~a compares the wave-function after scattering for the exact potential
\begin{equation}
\label{eq:potexact}
 V(x_1,x_2) =  V_{\rm c}(x_1)+  V_{\rm c}(x_2),
\end{equation}
the center of mass approximation
\begin{equation}
\label{eq:potapprox}
 V(x_1,x_2) \approx  2V_{\rm c}((x_1+x_2)/2)
\end{equation}
and the effective potential~\cite{WeissCastin09}
\begin{equation}
\label{eq:potapproxbetter}
 V(x_1,x_2) \simeq 2V_{\rm c}((x_1+x_2)/2) + \frac{V_{\rm c}''((x_1+x_2)/2)}{8\widetilde{\beta}^2}\;,
\end{equation}
where $\widetilde{\beta}$ is given by Eq.~(\ref{eq:betatilde}).
The effective potential clearly is a huge improvement over the center-of-mass approximation. It still remains to be shown that the final state indeed is entangled in the sense that whenever one measures one particle on the right (left) side, the other particle will be on the same side.To do this, a Measure If An Outgoing Wave is Unentangled, 
\begin{equation}
\label{eq:pmiaou}
 p_{\rm miaou}(t) \equiv \int_{-\infty}^0 d x_1\int^{-\infty}_0 d x_2 |\Psi(x_1,x_2;t)|^2,
\end{equation}
where the wavefunction is taken to be normalised to one. Figure~\ref{fig:miaou}~b demonstrates that for the wave-function to be entangled, the initial velocity cannot be too large. Experimental verifications that such states are indeed quantum superpositions  could be done via interference experiments~\cite{WeissCastin09}.

\section{\label{sec:conclusion}Conclusion}

To summarise, scattering of bright quantum matter wave
solitons~\cite{WeissCastin09,Streltsov-2008} has been investigated for two attractive atoms
in an optical lattice via exact numerics. It has been demonstrated that the beyond
centre-of-mass approximation approach of Ref.~\cite{WeissCastin09} is indeed necessary to
describe the physical situation at least qualitatively:  the effective potential derived in
Ref.~\cite{WeissCastin09} can explain  the break-down of the centre-of-mass approximation
observed for specific parameters in the numerics (Fig.~\ref{fig:three}).

For a scattering potential given by a laser focus (the potential would then be approximately Gaussian) such
an effect does not occur. The wavefunction then splits into two parts as shown for a single particle in
Fig.~\ref{fig:refl} and the effective potential even gives excellent quantitative agreement (Fig.~\ref{fig:miaou}) with exact numerics.

\acknowledgments

I acknowledge insightful discussions with Y.\ Castin, A.\ Sinatra and C.\ Salomon at the Laboratoire Kastler Brossel at the ENS in Paris where this research was started (funded by the
European Union via contract
 MEIF-CT-2006-038407). Furthermore, I would like to thank M.~Holthaus for his
continuous support.
\vspace*{4cm}

%\newpage

\end{document}